\documentclass[9pt,twocolumn,twoside]{osajnl}
\journal{ol}
\setboolean{shortarticle}{true}

\usepackage{overpic}
\usepackage{amsmath}
\usepackage{siunitx}
\DeclareMathOperator{\sech}{sech}

\usepackage[overlay,absolute]{textpos}
\newcommand\PlaceText[3]{%
	\begin{textblock*}{10in}(#1,#2)
		#3
	\end{textblock*}
}%

\title{Ultrafast mid-infrared fiber laser mode-locked using frequency-shifted feedback}

\author[1,*]{Matthew R. Majewski}
\author[1]{Robert I. Woodward}
\author[1]{Stuart D. Jackson}

\affil[1]{MQ Photonics, School of Engineering, Faculty of Science and Engineering, Macquarie University, North Ryde, NSW 2109, Australia}

\affil[*]{Corresponding author: matthew.majewski@mq.edu.au}

\begin{abstract}
We demonstrate ultrashort pulse generation from a fluoride fiber laser co-doped with holmium and praseodymium. 
To date the majority of work focused on short pulse generation from this class of fiber laser has employed loss modulators in the cavity, both real and artificial.
In this work we alternatively employ a frequency shifting element: an acousto-optic modulator (AOM) in the cavity.
This results in mode-locked output of sub-5~ps pulses with 10~nJ of energy at a center wavelength of 2.86~\si{\micro\meter}, and a pulse repetition frequency of 30.1~MHz, equating to a peak power of 1.9~kW.
Additional experimental investigation of the relationship between frequency shift and cavity round trip offer insight into the complex underlying dynamics.
As a complementary mode-locking technique to conventional loss modulation, this method of pulse-formation may greatly expand the design flexibility of pulsed mid-infrared fiber lasers. 
\end{abstract}

\setboolean{displaycopyright}{true}

\begin{document}

\maketitle

\PlaceText{25mm}{9mm}{Vol. 44, Issue 7, pp. 1698-1701 (2019); https://doi.org/10.1364/OL.44.001698}

Mode-locked fiber lasers in the mid-infrared (mid-IR) spectral region are promising sources for applications across an array of disciplines including medicine, manufacturing, and defense, due to the existence of strong molecular absorption features at mid-IR wavelengths.
While  a number of laser architectures are currently being considered for ultrashort mid-IR pulse generation, including nonlinear wavelength conversion of ultrafast near-IR sources and Cr/Fe-doped bulk crystal oscillators, there is particular demand for fiber-based systems, due to their compact design and flexibility.
This has led to significant sustained research effort to reliably generate short intense pulses from a variety of mid-IR fiber lasers~\cite{zhu2017pulsed}.

One particularly important design factor is the choice of mode-locking technique to modulate the cavity light.
The majority of mode-locked mid-IR fiber laser development has thus far focused on loss modulation via `real' saturable absorbers (SAs), i.e. materials exhibiting intensity-dependent absorption.
While SAs are widely available in the near-IR, the optimum choice of nonlinear materials for mid-IR SAs remains an open problem.
Recent works have demonstrated picosecond pulse generation using InAs semiconductors~\cite{hu2014stable, haboucha2014fiber} and also more exotic nanomaterials such as graphene~\cite{zhu2016graphene}, and black phosphorous~\cite{qin2016mid}.
However, these devices have yet to reach maturity as questions remain for semiconductors regarding their wavelength coverage (e.g. limited by the band edge) and response time (typically 10s ps, which limits output pulse durations~\cite{haboucha2014fiber}), and for nanomaterials regarding their long-term stability and additional undesirable nonlinear effects that can limit the modulation depth~\cite{malouf2018two}.
To date, the shortest SA-generated mid-IR pulse is 6~ps~\cite{hu2014stable}.

In terms of decreasing mid-IR pulse durations to the femtosecond level, `artificial' SAs based on nonlinear polarization evolution (NPE) have been successfully demonstrated.
Despite impressive results including the generation of sub-200~fs pulses from both Er- and Ho-doped fiber lasers around 2.8~\si{\micro\meter}~\cite{duval2015femtosecond,antipov2016high}, NPE cavities are intrinsically sensitive to environmental variations, precluding their deployment for many practical applications.
There is therefore still strong demand for robust mode-locking techniques that can be simply deployed in the mid-IR.

One such method, dating back to early laser designs yet relatively understudied for modern systems, is to place a frequency shifting element in the cavity, which produces pulses through a novel frequency-shifted feedback (FSF) dynamic, rather than relying on direct loss modulation.
Specifically, the FSF arrangement disrupts the classical longitudinal mode structure of the laser by monotonically shifting the frequency (wavelength) of the signal light each round trip, eventually pushing the signal beyond the bandwidth of either an explicitly inserted wavelength selective filter, or the gain bandwidth itself.
In a continuous wave (CW) sense this results in emission which is essentially incoherent amplified spontaneous emission~\cite{littler1991continuous}.

This arrangement has also been shown to generate short pulses, however, when there is sufficient nonlinearity in the cavity (e.g. fiber Kerr nonlinearity).
Several works have confirmed the pulse forming dynamics via detailed numerical models~\cite{de1995simple, sabert1994pulse,woodward2018mode}, but in brief, the pulse generation dynamic can be explained by noting that when nonlinearity is included, intense light experiences spectral broadening through self-phase modulation.
This works to counteract the continuous frequency shifting out of the gain bandwidth as intense light broadens a portion of its energy back towards the center of the filter, thus providing preferentially higher gain to short intense pulses than for low-intensity CW light.
Consequently, the laser operates in a pulsed steady state, generating a stable pulse train at the cavity round-trip frequency.
While this is not `mode-locking' by the rigorous definition of phase-locked longitudinal modes, the resultant pulsed output shares many commonalities with mode-locking, so in keeping with literature on the topic~\cite{sabert1994pulse} we adopt the term here.

FSF pulse generation has been successfully demonstrated in near-IR fiber systems~\cite{sousa2000short}, though has now been largely abandoned in favor of semiconductor SAs for practical systems due to high-performance commercially available near-IR SAs.
At longer wavelengths, however, where robust SAs are not currently available yet acousto-optic frequency shifters are well developed (e.g. based on TeO$_2$, suitable for operation from 2.0 to 4.5~\si{\micro\meter}), this FSF method holds great promise.
We have recently demonstrated FSF mode-locking for the first time in the mid-IR using a dysprosium-doped ZBLAN gain fiber with an intracavity acousto-optic tunable filter (AOTF), achieving 33~ps pulses and record tunability from 2.97 to 3.30~~\si{\micro\meter}~\cite{woodward2018mode}. 
In this letter we report on the extension of the FSF method to a Ho$^{3+}$:ZBLAN system which replaces the AOTF with an acousto-optic modulator (AOM) and enables the generation of 4.7~ps pulses---an order of magnitude shorter.
The underlying FSF dynamics that enable such an improvement are discussed, and investigations showing how the relationship between frequency shift magnitude and cavity free spectral range affects pulse stability are presented, offering new insight into the FSF technique.

\begin{figure}[tbp]
\includegraphics{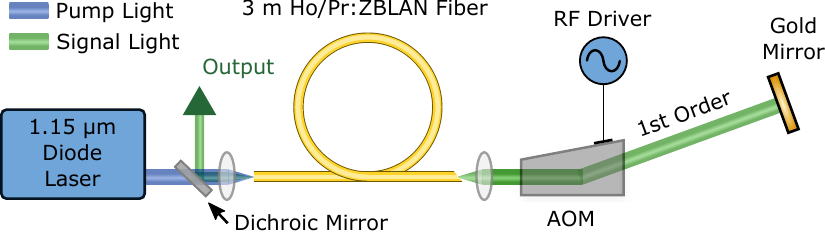}
\caption{Experimental layout of the frequency-shifted feedback fiber laser.}
\label{fig:layout}
\end{figure}

The experimental setup for our FSF laser is shown in Fig.~\ref{fig:layout}.
The fiber is 3~m of double-clad co-doped Ho$^{3+}$/Pr$^{3+}$:ZBLAN (1000/20,000 mol. ppm respectively) with an outer cladding diameter of \SI{125}{\micro\meter}, a core diameter of \SI{15}{\micro\meter} and a 0.13 core NA.
Co-doping in this case serves to de-populate the Ho lower laser level, allowing for efficient diode pumping~\cite{jackson2009high}, which we provide with 1150~nm multimode laser diodes.
A perpendicular cleave on the pump injection end of the fiber serves as the laser output coupler, while pump and signal are separated by a \SI{45}{\degree} dichroic mirror.
The distal end of the fiber is angle cleaved to avoid parasitic lasing.
Emission is collimated by an aspheric ZnSe lens before passing through the AOM (Gooch \& Housego) which is driven by a variable RF sinusoid, initially set at 31~MHz.
Finally, the cavity is closed by reflecting the first-order diffracted light (diffracted with 70\% efficiency) with a gold mirror.
The TeO$_2$ crystal AOM is a traveling acoustic wave design which imparts a frequency shift to the diffracted cavity light equal to the applied RF drive.
As the AOM is driven continuously, it operates in the cavity solely as a frequency shifter rather than a modulator which is switched on and off.

\begin{figure}[tbp]
	\centering
	\includegraphics{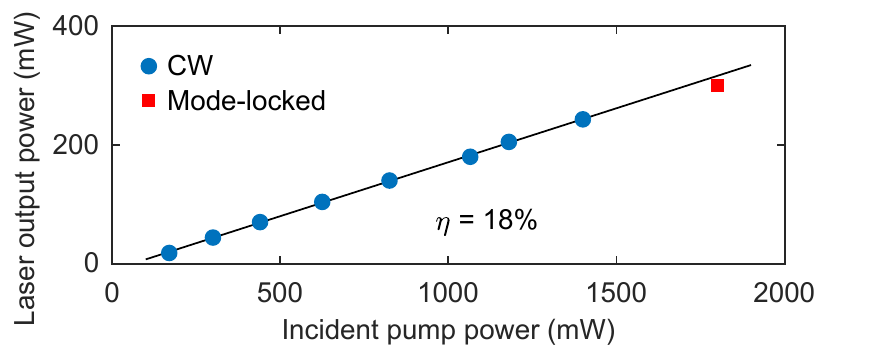}
	\caption{Laser output power in continuous wave (CW) and mode-locking operation; slope efficiency ($\eta$) is nominally 18\% with respect to incident pump power.}
	\label{fig:slope}
\end{figure}

At 60~mW incident pump threshold, lasing is observed, free-running at 2.864~\si{\micro\meter}.
Laser output power as a function of incident pump power is shown in Fig.~\ref{fig:slope}, exhibiting a slope efficiency of 18\%.
At low pump powers, before the onset of mode-locking, it should be noted that in this CW state the output is temporally noisy, but exhibits no strong discernible structure, when recorded with a photodiode and oscilloscope.
This could be explained by the frequency shifted feedback continually disturbing the cavity longitudinal structure, but with insufficient intracavity power to initiate stable pulsation---such disturbances prevent relaxation oscillations being fully damped to reach a coherent constant power steady state~\cite{sabert1994pulse}.
At pump powers exceeding 1.5~W, mode-locking can be initiated, typically by applying a minor mechanical perturbation to the cavity.
The pulse repetition rate of 30.1~MHz corresponds to the inverse cavity round-trip time ($f_\mathrm{0}$).

\begin{figure}[tbp]
	\centering
	\begin{overpic}{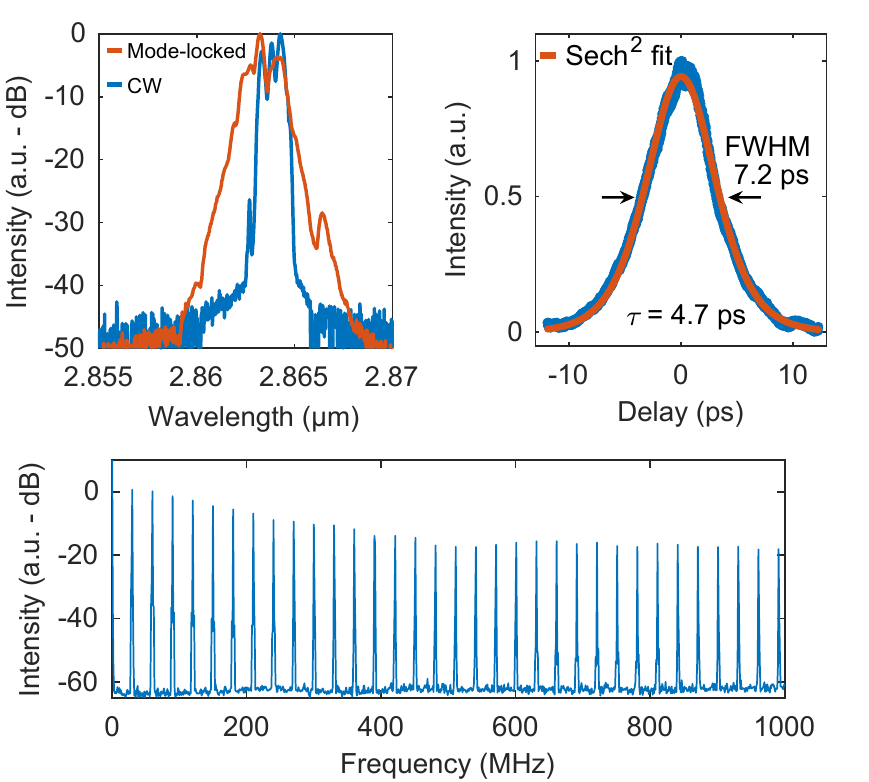}
	\put(0,85){(a)}
	\put(50,85){(b)}
	\put(0,40){(c)}
	\end{overpic}
	\caption{Mode-locked laser output: a)~optical spectrum exhibiting broadening beyond CW linewidth. b)~measured autocorrelation trace and $\sech^2$ fit with a full width half-max (FWHM) width of 7.2~ps, deconvolved to a pulse width ($\tau$) of \SI{4.7}{\pico\s}. c)~RF spectrum.}
	\label{fig:mldata}
	
\end{figure}

While in a mode-locked state, the laser output is characterized both spectrally and temporally. 
The optical spectrum measurably broadens during mode-locked operation as compared to the CW spectrum, as seen in Fig.~\ref{fig:mldata}(a).
The pulse duration is measured via an in-house constructed two-photon absorption autocorrelator, and as seen in Fig.~\ref{fig:mldata}(b) the measured trace is well fitted by a sech$^2$ pulse with a deconvolved width of 4.7~ps.
This pulse duration compares favorably to the previous shortest pulse from a non-NPE system of 6~ps~\cite{hu2014stable}.
Additionally, this duration is almost an order of magnitude shorter than our previous mid-IR FSF laser using Dy-doped fiber~\cite{woodward2018mode}.
This is attributed to the broader effective filter bandwidth in the present laser, which has been theoretically shown to yield shorter pulses through FSF dynamics~\cite{de1995simple}.
Whereas in Ref.~\cite{woodward2018mode} the inclusion of an AOTF provided simultaneous frequency shifting and spectral filtering with 5~nm bandwidth, in our system here, we use an AOM to generate the frequency shift, which for a fixed angular alignment can also act as a spectral filter.
This can be explained by noting that diffraction efficiency is sensitive to deviations from the Bragg angle, which is explicitly defined for monochromatic light.
Thus the AOM effectively becomes increasingly `mis-aligned' (less efficient diffraction) as the wavelength deviates from the Bragg angle definition.
Spectral bandwidth of this effect can be reasonably estimated from the physical properties of the AOM i.e. drive frequency, acoustic velocity, and crystal dimensions, from which we estimate in this case a bandwidth of 10s nm.

At 1.8~W pump power, the maximum output average power is 300~mW, corresponding to 10~nJ pulse energy and 1.9~kW peak power.
Attempts to further increase the power resulted in fiber facet damage, requiring the end to be re-cleaved.
This is a widely acknowledged problem with ZBLAN fibers, owing to their inferior thermo-mechanical durability compared to silica glass.
However, recent works have demonstrated that end-capping ZBLAN fiber with more a robust fluoride glass (e.g. AlF$_3$) can significantly improve the power-handling~\cite{aydin2018towards}.
Thus we believe that our FSF laser design could support even higher pulse energies with appropriate fiber processing, which is a topic of ongoing work.

To evaluate the pulse train stability, photodiode measurement of the output in the radio frequency (RF) domain is shown in Fig.~\ref{fig:mldata}(c), with harmonics of the characteristic roundtrip frequency (30.1~MHz) out to 1~GHz.
The steady train of harmonics shows the absence of undesired long-term envelope modulation, which is a strong indicator of stability. 

\begin{figure}[!h]
	\centering
	\includegraphics{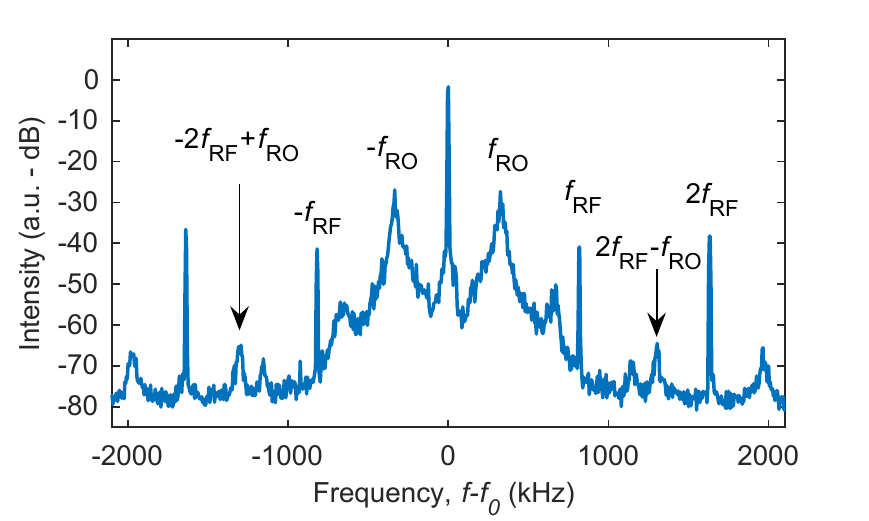}
	\caption{RF spectrum of the fundamental beat note corresponding to the inverse of cavity round trip time (\textit{f}$_0=$~\SI{30.1}{\mega\hertz}). Relaxation oscillations (RO), AOM RF drive frequency, and the mixture of both are seen.}
	\label{fig:RFfundamental}
\end{figure}

In order to more fully investigate the temporal stability of our pulse train, we acquire high resolution RF data around the fundamental beat note $f_0$ as seen in Fig.~\ref{fig:RFfundamental}.
Additional peaks at a frequency $f_{\mathrm{RO}}$ of 320~kHz from $f_0$ are observed, approximately 30-dB lower in intensity than the fundamental beat, which are attributed to weak relaxation oscillations. 
The RF drive frequency of the AOM ($f_{\mathrm{RF}}$) is also a visible feature.
Finally, it is particularly interesting to note that there also appears to be mixing of $f_{\mathrm{RF}}$ and $f_{\mathrm{RO}}$, producing additional peaks in the RF spectrum.

\begin{figure}[!hb]
	\centering
	\begin{overpic}{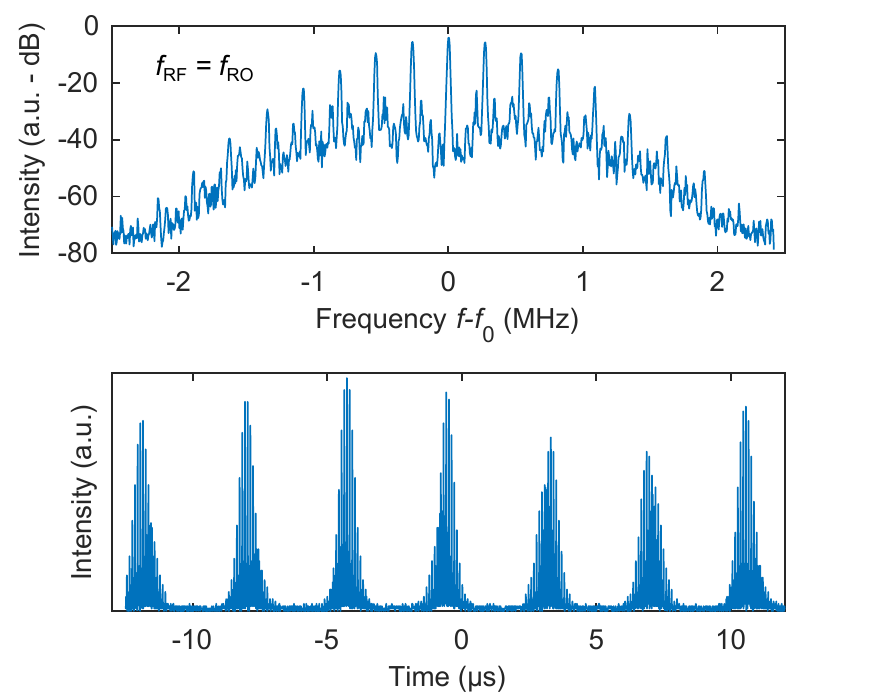}
	\put(5,0){(b)}
	\put(5,43){(a)}	
	\end{overpic}
	\caption{Q-switched mode-locking resulting from matching RF drive to characteristic relaxation oscillation frequency ($f_{\mathrm{RF}}$~=~$f_{\mathrm{RO}}$); a) RF spectrum of fundamental. b) oscilloscope trace.} 
	\label{fig:qs_ml}
\end{figure}

This observation motivated us to further investigate the dynamic interplay between frequency shift and cavity free-spectral range (which to our knowledge, has yet to be studied), by adjusting the applied drive frequency to the AOM, which sets the acoustic transducer frequency and thus defines the frequency shift of diffracted light.
By placing the AOM drive frequency closely matched to the characteristic relaxation oscillation frequency $f_{\mathrm{RO}}$, a mode-locked state is completely inhibited.
Instead, the relaxation oscillations appear to be strongly reinforced, as seen in the RF data presented in Fig.~\ref*{fig:qs_ml}(a).
Here we see many relaxation oscillation harmonics, as well as a broad RF pedestal associated with large uncertainty in pulse energy~\cite{von1986characterization}.
In this case, a temporal domain oscilloscope trace of the same output state in Fig.~\ref*{fig:qs_ml}(b) shows Q-switched mode-locking (i.e. $\sim$1~\si{\micro\second} bursts of mode-locked pulses related to a dynamical instability from undamped relaxation oscillations).

\begin{figure}[tbp]
	\centering
	\begin{overpic}{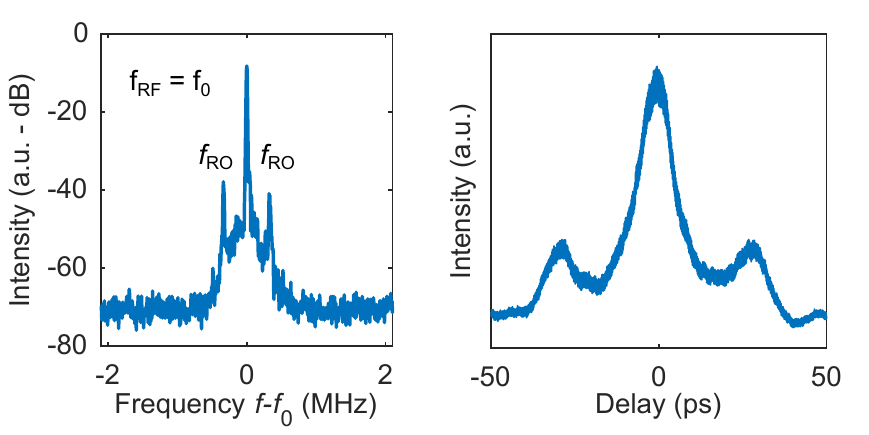}
		\put(5,0){(a)}
		\put(50,0){(b)}	
	\end{overpic}
	\caption{Multiple pulse operation when $f_{\mathrm{RF}}$~=~$f_0$~: a) RF spectrum of fundamental, b) autocorrelation trace.}
	\label{fig:mp_ml}
\end{figure}

Furthermore, another distinct operation regime is encountered when the AOM drive frequency is tuned to match the cavity round trip time ($f_{\mathrm{RF}}$ = $f_0$).
An RF trace of the fundamental beat in this situation is presented in Fig.~\ref{fig:mp_ml}(a).
In comparison to previous cases, the relaxation oscillation peaks are increasingly suppressed and the central pedestal is reduced, indicating a higher degree of temporal stability.
While a steady mode-locked pulse train is observed on the oscilloscope, the autocorrelation trace acquired in this regime (Fig.~\ref{fig:mp_ml}(b)) indicates clearly multi-pulsing.
The symmetric nature of autocorrelation measurements prevents a detailed description of the pulse shape, but this trace is characteristic of a main pulse, followed by a weaker `satellite pulse'. 
 
Based on the current understanding of FSF mode-locked system, there is not a direct physical relationship between the frequency shift (here $f_{\mathrm{RF}}$) and the inverse round trip time ($f_0$) that is required to obtain stable pulse generation.
Indeed, FSF mode-locking in the near-IR has previously been shown for a fixed-length fiber cavity with frequency shifts varying from 25~kHz to 100~MHz~\cite{sabert1994pulse}.
However, the data presented in Fig.~\ref{fig:RFfundamental} shows a clear dependence of the pulsed operating state on the ratio of frequency shift to cavity round-trip frequency when the two values are within close proximity.
To our knowledge, this effect has not previously been observed in FSF mode-locked lasers. 

While the above behavior is at present not completely understood, it is possible that when $f_\mathrm{0}\sim f_\mathrm{RF}$, there are contributions to pulsation from both FSF and frequency modulation (FM) mode-locking dynamics.
In FM modelocking, which is perhaps more accurately described as phase modulation mode-locking, a periodic phase modulation matched to the cavity round trip time can produce pulses~\cite{kuizenga1970fm}.
The discussion of this system thus far has focused on the intentional (and well defined) frequency shift imparted by the AOM, however the acoustic waves of the modulator also impart some phase shift, which is essentially arbitrary.
In contrast, as previously discussed, FSF modelocking is a wholly separate process relying on fundamentally different dynamics.
It can be argued that this particular system is circumstantially quite close to an FM modelocked system, while retaining the frequency shift of a FSF system, limiting the pure stability of either.
Possible further experimentation to separate out these potential effects could focus on substantial alteration of cavity roundtrip while attempting to maintain total cavity gain.
An alternatively designed AOM with different diffraction bandwidth may also elucidate the underlying dynamics.

Therefore, a detailed experimental and numerical study of these dynamics is a valuable topic for further work, which will improve understanding of the FSF dynamics, and could lead to improved temporal stability for FSF mode-locked sources.
An additional interesting direction for future work is to deploy the FSF technique for pulse generation at even longer wavelengths using other transitions in doped fibers, e.g. 3.5~\si{\micro}m Er$^{3+}$~\cite{henderson2016versatile}, 3.9~\si{\micro}m Ho$^{3+}$~\cite{maes2018room}, or >4~\si{\micro}m Dy$^{3+}$~\cite{majewski2018emission}.
FSF is uniquely well-suited here as TeO$_2$ AOMs are already commercially available for operation up to 4.5~\si{\micro\meter}.

In conclusion we have demonstrated the first ultrafast FSF mode-locked mid infrared fiber laser, using a simple linear cavity including an AOM frequency shifter.
Stable pulses with 4.7~ps duration were achieved at 2.86~~\si{\micro\meter}, with up to 10~nJ, shorter than the shortest pulses produced from saturable absorber-based mid-IR fiber lasers to date.
Evidence of previously unreported complex underlying dynamics were also observed, including a dependence of the operating state on the relationship between frequency shift and round-trip frequency, motivating further work to more completely understand these effects.
Certainly, the FSF technique is a promising direction for the generation of ultrashort pulses from mid-IR fiber lasers.

\section*{Funding Information}
Australian Research Council (ARC) (DP170100531).
RIW acknowledges support through an MQ Research Fellowship.

\end{document}